\documentclass{INTERSPEECH2023}

\interspeechcameraready

\usepackage{colortbl}
\usepackage{tcolorbox}

\def\tabref#1{Table \ref{#1}}
\def\figref#1{Figure \ref{#1}}
\def\secref#1{Section \ref{#1}}
\def\eqref#1{Eq (\ref{#1})}

\definecolor{deepgreen}{rgb}{ .0, .44, .0} 

\title{On the Audio Hallucinations in Large Audio-Video Language Models}
\name{Taichi Nishimura, Shota Nakada, Masayoshi Kondo}
\address{
  LY Corporation
\email{\{tainishi,shota.nakada,masayoshi.kondo\}@lycorp.co.jp}
}
\begin{document}

\maketitle
 
\begin{abstract}
Large audio-video language models can generate descriptions for both video and audio. However, they sometimes ignore audio content, producing audio descriptions solely reliant on visual information. This paper refers to this as audio hallucinations and analyzes them in large audio-video language models.
We gather 1,000 sentences by inquiring about audio information and annotate them whether they contain hallucinations. If a sentence is hallucinated, we also categorize the type of hallucination. The results reveal that 332 sentences are hallucinated with distinct trends observed in nouns and verbs for each hallucination type.
Based on this, we tackle a task of audio hallucination classification using pre-trained audio-text models in the zero-shot and fine-tuning settings. Our experimental results reveal that the zero-shot models achieve higher performance (52.2\% in F1) than the random (40.3\%) and the fine-tuning models achieve 87.9\%, outperforming the zero-shot models.
\end{abstract}

\section{Introduction}

Towards perfect video understanding, learning videos with audio is essential because audio offers complementary information to the visual modality.
One notable task in this pursuit is audio-video captioning \cite{Shen_2023_CVPR,chen2023valor,Tian_2019_CVPR_Workshops,liu23l_interspeech}, which aims to describe both visual and audio content in natural language.
While supervised approaches have demonstrated robust performance on benchmark datasets, their effectiveness is constrained to the training dataset. In other words, they struggle to generalize well to videos that are out-of-domain from the training set (e.g., the models trained from vlog fail to describe cartoon videos).

\begin{figure}[t]
  \centering
  \includegraphics[width=\linewidth]{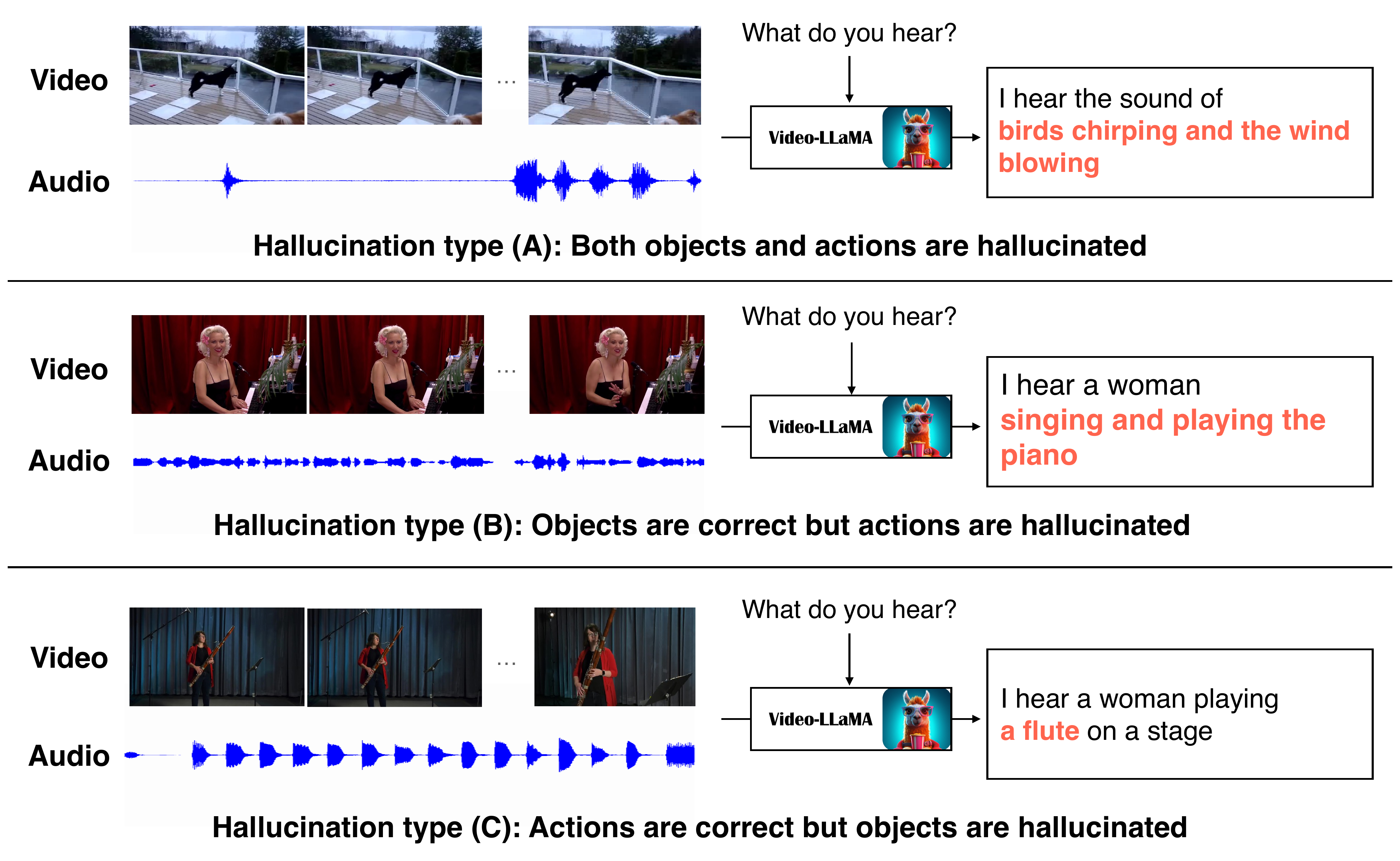}
  \caption{Examples of audio hallucinations. We collect 1,000 response sentences from Video LLAMA and annotate whether the sentences are hallucinated or not. If a sentence is hallucinated, we annotate the hallucination types with it.}
  \label{fig:task_overview}
\end{figure}

\begin{figure*}[t]
  \centering
  \includegraphics[width=0.9\linewidth]{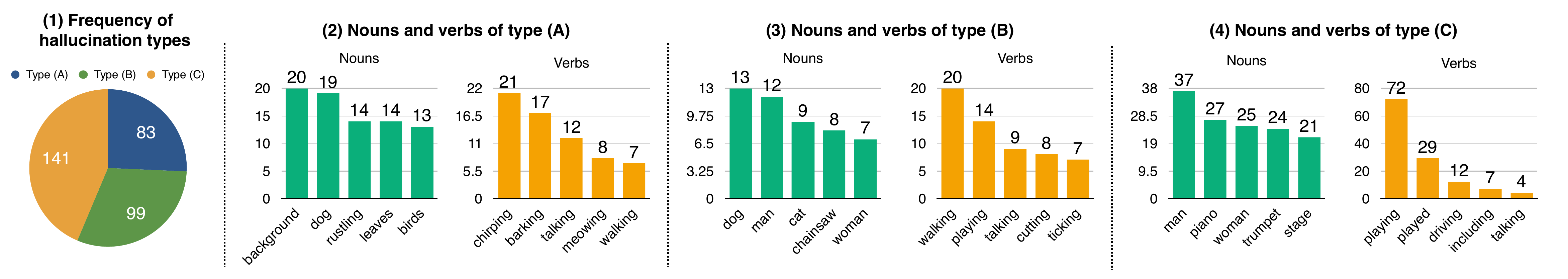}
  \caption{Statistics on the hallucination types.}
  \label{fig:hallucination_types}
\end{figure*}

\begin{figure*}[t]
  \centering
  \includegraphics[width=0.9\linewidth]{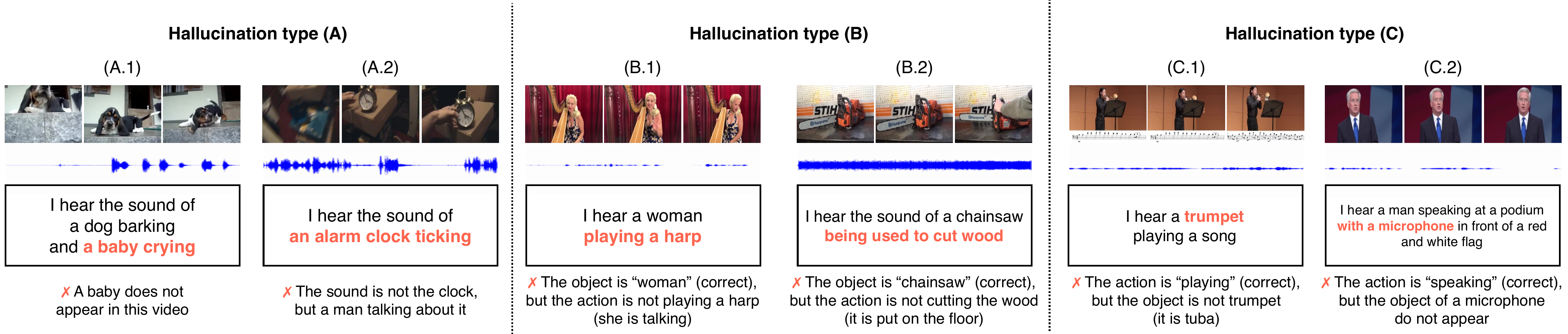}
  \caption{Audio hallucination examples for each type.}
  \label{fig:hallucination_examples}
\end{figure*}

Recently, significant attention has been drawn to large language models (LLMs) like GPT-4 \cite{openai2023gpt4} and LLAMA2 \cite{touvron2023llama}, owing to their robust generalization capabilities. Researchers have extended the scope of LLMs into the multimodal domain, exploring image-text, audio-text, and video-text LLMs \cite{pmlr-v202-li23q,deshmukh2023pengi,li2024videochat}.
In line with these studies, Video LLAMA, a video-audio language model, has been proposed \cite{zhang-etal-2023-video}. Video LLAMA exhibits the capability to generate descriptive content for both video and audio, showcasing impressive video understanding skills in conveying the essence of their respective contents.

However, we observe that Video LLAMA sometimes overlooks audio content, producing audio descriptions solely reliant on visual information (\figref{fig:task_overview}).
We refer to this behavior as \textbf{audio hallucinations} as discussed in the context of generating unfaithful descriptions in previous LLM studies. The audio hallucination is categorized into three types: (A) both objects and actions are hallucinated, (B) objects are correct but the actions are hallucinated, and (C) the actions are correct, but the objects are hallucinated.

This paper focuses on analyzing the audio hallucinations of large-scale audio-video language models. The analysis comprises four key processes.
Firstly, we formulate prompts to inquire about audio information and gather 1,000 responses from the Video LLAMA. Secondly, we manually annotate whether the sentences are hallucinated or not. Thirdly, we also annotate the hallucination type associated with the hallucinated sentences.
Finally, we aggregate the annotation results, revealing that 323 sentences were indeed hallucinated, with distinct trends observed in nouns and verbs for each hallucination type.
It is noteworthy that the hallucination rates of state-of-the-art LLMs stand at a mere 2.0\%.
In contrast, our findings show Video LLAMA has a substantially higher audio hallucination rate of 32.3\%, underscoring a significant departure from the performance of these advanced LLMs.

Based on the annotations, we tackle the task of audio hallucination classification in the zero-shot and fine-tuning settings. Because the audio hallucinations are caused by relying on visual information, we propose a method to classify them by using pre-trained audio-text models. Our experimental results demonstrate that the zero-shot methods achieve higher performance (52.1\%) than the random baseline (40\%) and the fine-tuning methods achieve 87.9\% in F1, outperforming the zero-shot methods. Qualitative evaluation reveals two insights. First, the zero-shot model correctly predicts the single sound of animals but fails to predict the mixed sounds and less frequent instruments. Second, the fine-tuning model answers such sounds correctly but still fails to predict the confusing sound based on only the sound, such as a running tractor and lawnmower.

\section{Related work}

\subsection{Audio captioning}

Audio captioning is a task to generate natural language descriptions from the audio \cite{kim-etal-2019-audiocaps,mei:2022:icassp,mei:2021:dcase}.
As with image captioning \cite{Vinyals_2015_CVPR}, previous approaches employ an encoder-decoder architecture \cite{NIPS2014_a14ac55a}, which connects the convolutional neural networks (CNNs) and recurrent neural networks (RNNs). With a rapid advance of Transformers \cite{NIPS2017_3f5ee243}, Transformer-based audio captioning models have been proposed \cite{mei:2021:dcase} and achieved strong ability on benchmark datasets, such as Clotho \cite{9052990} and AudioCaps \cite{kim-etal-2019-audiocaps}.

The key limitation of audio captioning is that there exists indistinguishable sound only from the audio information, such as the sound of a working tractor and lawnmower. Video-audio captioning is one of the important topics in audio-visual learning \cite{10.1007/978-3-030-01216-8_16,Sun_2023_CVPR,Chen_2023_CVPR,montesinos2022vovit,Zhang_2022_CVPR}, aiming to generate the correct descriptions by fusing audio and visual information effectively.
Video LLAMA demonstrated strong capabilities of understanding audio information in the videos. However, we observe that it sometimes ignores the audio information to generate audio descriptions. This paper is the first attempt to define this behavior as audio hallucinations and analyze them for details.

\subsection{Hallucinations in LLMs}

Although LLMs demonstrated remarkable abilities in producing fluent and accurate responses based on input text prompts, they still suffer from hallucinations, where they generate misleading or unfaithful descriptions fluently.
Detecting hallucinations has become a prominent topic in natural language processing, and researchers rapidly have developed benchmark datasets and approaches recently \cite{manakul-etal-2023-selfcheckgpt,liu-etal-2022-token,durmus-etal-2020-feqa}.

Although previous approaches focused on detecting hallucinations in single or dual modalities (e.g., vision and language \cite{anonymous2023analyzing}), no work investigated the hallucinations in triple modalities. This study is the initial work to scrutinize audio hallucinations in video-audio language models and introduces a method for their detection in both zero-shot and fine-tuning settings.

\section{Audio hallucination analysis}
\label{sec:audio_hallucination_analysis}

In this section, we describe how to construct a corpus to analyze the audio hallucination of large audio-video language models and report the analyzed results. Our analysis consists of three processes: (1) annotating audio hallucination with the generated sentences, (2) investigating the statistics such as frequent nouns and verbs, and (3) investigating the hallucinated examples.

\subsection{Annotation}
\label{subsec:audio_hallucination_annotation}

As the audio-video language model, we employ Video-LLAMA-7B \cite{zhang-etal-2023-video}, which is the state-of-the-art audio-video LLMs and can describe the content of video with audio in natural language. As the video-audio dataset, we use FAVDBench \cite{Shen_2023_CVPR}, which was recently introduced to evaluate the model's capabilities of describing both video and audio contents in natural language. FAVDBench contains 7,500/1,000/1,000 training/validation/testing samples, but we use only the test split.

We input the videos in the test set to Video-LLAMA. In this study, we focus on the capabilities of audio understanding of Video-LLAMA, thus we set a user prompt as \textit{What do you hear?}, which is officially used in the Video-LLAMA paper. Then, we collect the system outputs and conduct two kinds of annotations: (1) audio hallucination labels and (2) audio hallucination types.

The audio hallucination labels represent whether the target sentences are hallucinated or not. They are annotated in binary format and reveal how many sentences are hallucinated in the Video-LLAMA responses. In addition, to further analyze the audio hallucination, we annotate their hallucination types. We observe that audio hallucination can be categorized into three types: (A) both objects and actions are hallucinated, (B) objects are correct but actions are hallucinated, and (C) actions are correct but objects are hallucinated.

\subsection{Statistics on the constructed dataset}
\label{subsec:stats_audio_hallucination}

\figref{fig:hallucination_types} shows the statistics of the frequency of hallucination types and top-5 frequent types of nouns and verbs in each type.
To extract the nouns and verbs, we first remove commonly generated phrases of ``I hear,'' ``I hear that,'' and ``I hear the sound of'' and employ spaCy \footnote{\url{https://spacy.io/}} to them.

From the 1,000 sentences, we find that 323 sentences are hallucinated in total, indicating that Video-LLAMA suffers from audio hallucination. The type (C) is the most frequent among the hallucination types. In addition, we observe that frequent nouns and verbs are different between the hallucination types. For example, in type (A), the ``background'' is the most frequent noun, implying that the Video LLAMA hallucinates the sounds that do not appear in the video. In type (C), ``playing'' and ``played'' are the most frequent verbs, suggesting that the musical instruments are misrecognized from the audio.

\subsection{Qualitative evaluation}

\begin{figure}[t]
  \centering
  \includegraphics[width=\linewidth]{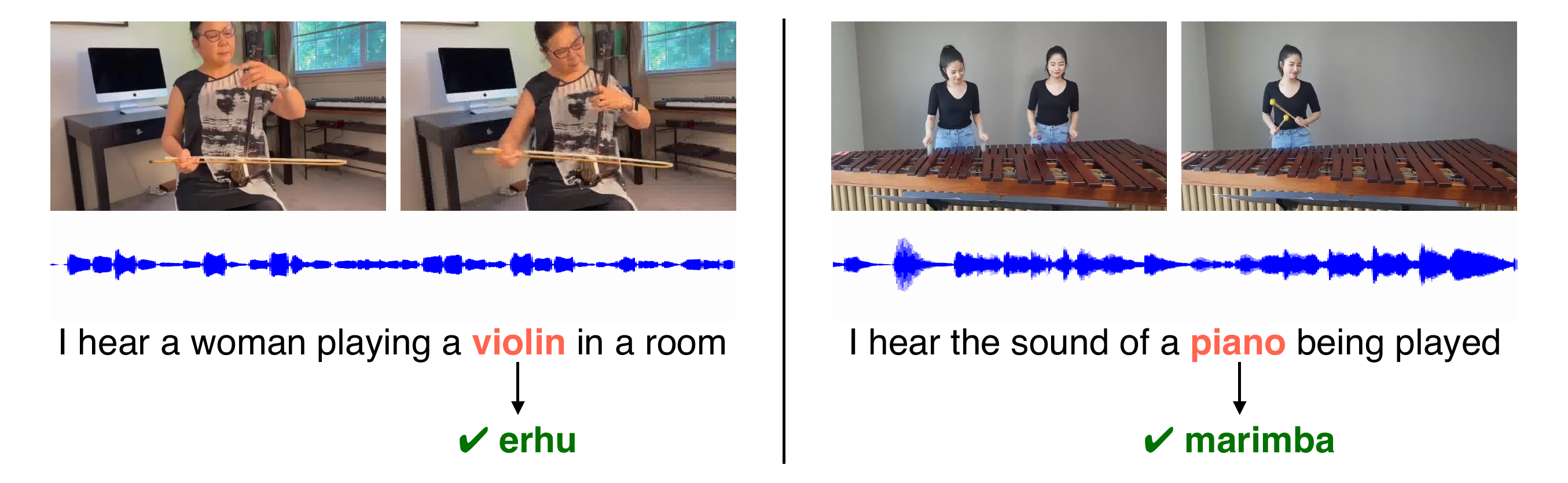}
  \caption{Audio hallucination cases on musical instruments.}
  \label{fig:music_instruments}
\end{figure}

\figref{fig:hallucination_examples} demonstrate the hallucination examples for each type. From these examples, we assume that these hallucinations occur because Video LLAMA ignores audio cues while placing a high emphasis on visual information. For example, in (A.1), ``a baby crying'' is generated but no baby appears in the video. In (B.1) the object of ``a woman'' is correct but ``playing a harp'' is incorrect, and in (C.1) the action of ``playing'' is correct but ``trumpet'' is incorrect. Therefore, to generate faithful responses, it is essential to strengthen the audio-text connections.

As discussed in \secref{subsec:stats_audio_hallucination}, type (C) hallucinations are due to the misrecognition of musical instruments. In (C.1), ``tuba'' is correct but ``trumpet'' is generated in failure.
\figref{fig:music_instruments} shows additional hallucination examples of such musical instrument cases. In many cases, less-frequent instruments are placed into frequent ones, such as piano, violin, and drum. Covering many kinds of instruments is essential for the future development of enhancing the audio understanding of multimodal LLMs.

\section{Audio hallucination classification}

\begin{figure}[t]
  \centering
  \includegraphics[width=\linewidth]{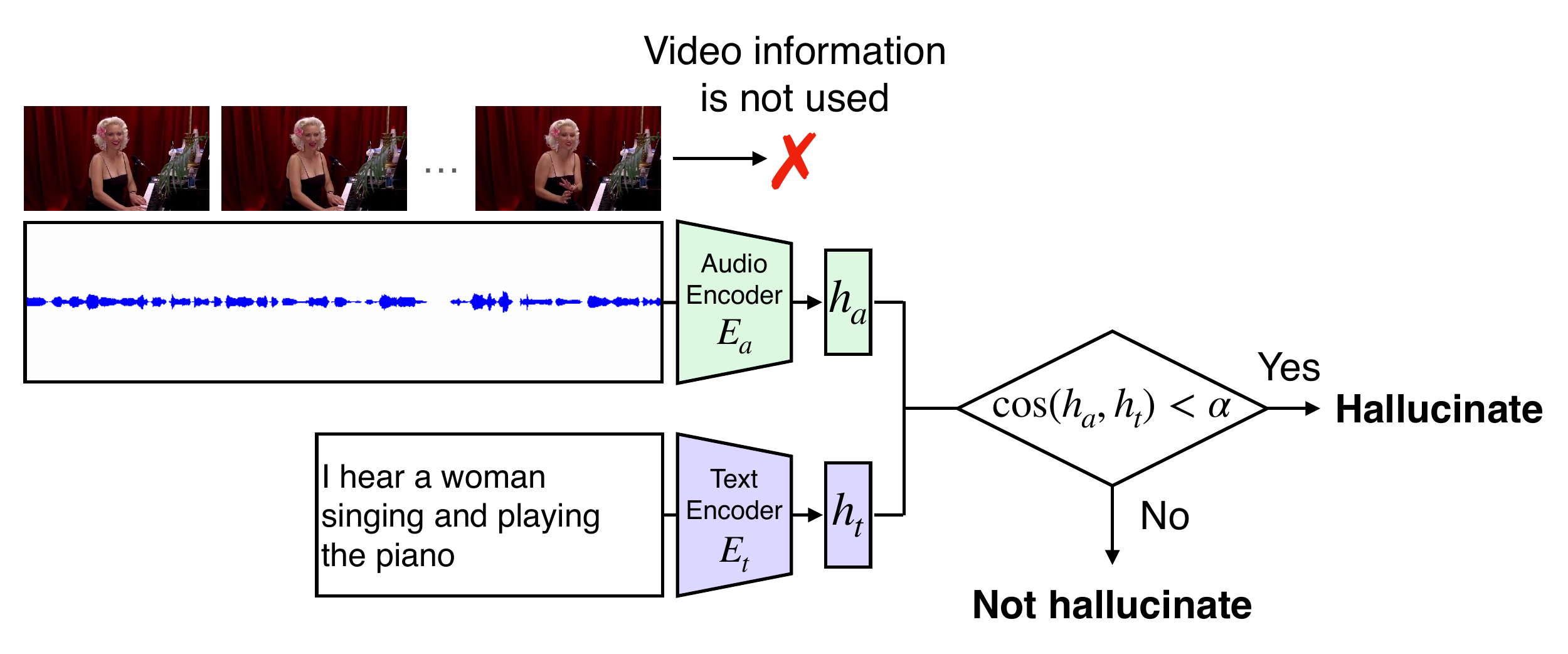}
  \caption{Zero-shot audio hallucination classifier.}
  \label{fig:classifier}
\end{figure}

\begin{figure}[t]
  \centering
  \includegraphics[width=\linewidth]{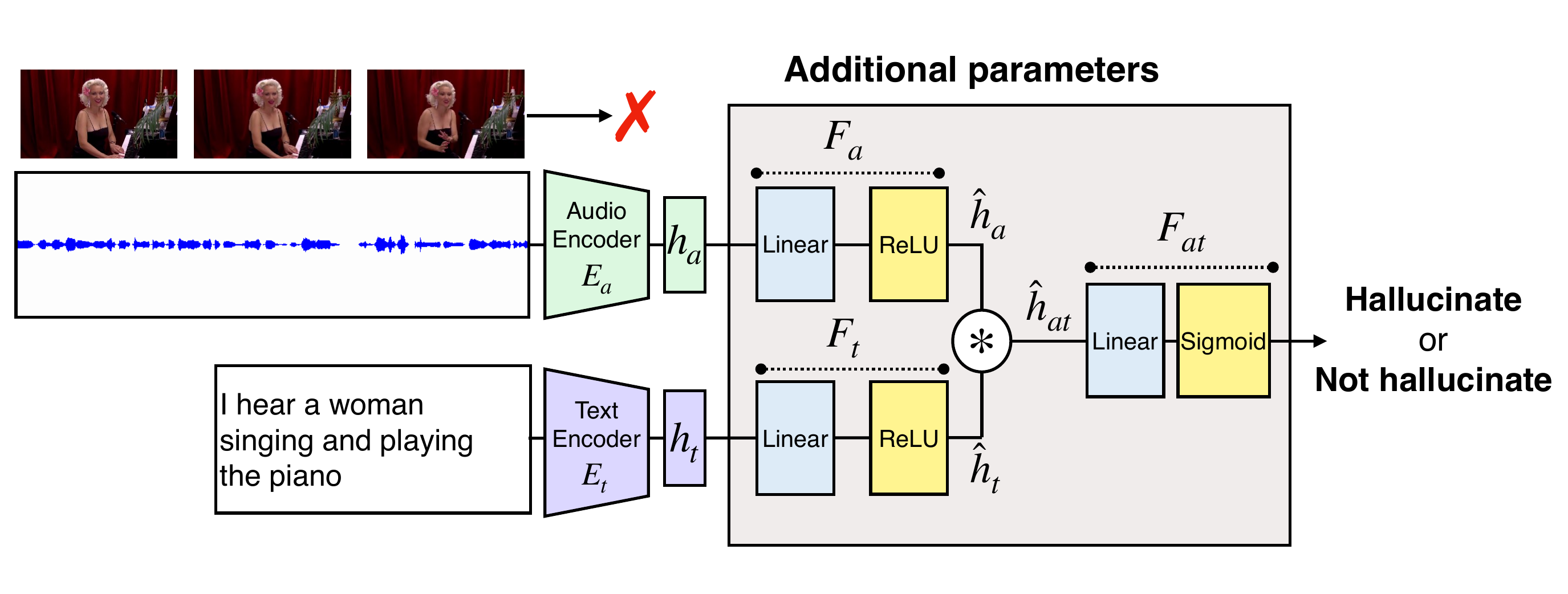}
  \caption{Fine-tuning audio hallucination classifier.}
  \label{fig:fine_tuned_classifier}
\end{figure}

In \secref{sec:audio_hallucination_analysis}, we observe that the audio hallucination occurs due to ignoring the audio information of the video.
Then, one question arises: \textit{can we detect the audio hallucination based on a pre-trained audio-text model?} We assume that the pre-trained audio-text models, such as MS-CLAP \cite{CLAP2023} and LAION-CLAP \cite{laionclap2023}, are better aware of connecting audio-text representations because they do not have a bypass of video-text connections.

Based on this assumption, we tackle the audio hallucination classification task using pre-trained audio-text models. We formulate this task as a binary classification and conduct experiments in the zero-shot and fine-tuning settings.

\subsection{Zero-shot approach}

\figref{fig:classifier} shows our zero-shot approach. Given audio $\mathbf{A}$ and text $\mathbf{T}$, the audio-text models convert them into a joint embedding space $\mathbf{h}_a = E_a(\mathbf{A}),\ \mathbf{h}_t = E_t(\mathbf{T})$, where $E_a,E_t$ represent the audio and text encoders and $\mathbf{h}_a, \mathbf{h}_t \in \mathbb{R}^d$ represent the embedded audio and text vectors.
Then, their cosine similarity $\cos(\mathbf{h}_a,\mathbf{h}_t)$ is computed as a score to output the label. The label of ``hallucinated'' is output if the score is smaller than $\alpha$, otherwise ``not hallucinated.'' Note that $\alpha$ is a hyperparameter.

\subsection{Fine-tuning approach}

\figref{fig:fine_tuned_classifier} shows the fine-tuning approach. We incorporate additional multi-layer perceptrons (MLPs) into the audio and text encoders. Let $F_a,F_t$ be audio and text MLPs (one linear layer + ReLU activation function). Given the audio and text vectors $\mathbf{h}_a,\mathbf{h}_t$, the output probability $y$ is computed as:
\begin{eqnarray}
\hat{\mathbf{h}}_a = F_a(\mathbf{h}_a), \ \ \hat{\mathbf{h}}_t = F_t(\mathbf{h}_t), \\
\hat{\mathbf{h}}_{at} = \hat{\mathbf{h}}_a \odot \hat{\mathbf{h}}_t,\ \ \hat{y}=F_{at}(\hat{\mathbf{h}}_{at}),
\end{eqnarray}
where $F_{at}$ represents the linear layer with sigmoid function and $\hat{y}$ is the predicted score. Based on the $\hat{y}$, we compute the binary cross entropy $L=BCE(\hat{y},y)$ as the loss function, where $y$ is the ground-truth label. Note that the audio and text encoders $E_a,E_t$ are frozen during the training.

\section{Experiments}

\begin{table}[t]
\centering
\caption{Hallucination classification results.}
\begin{tabular}{lccc}
\hline
 & Recall & Precision & F1 \\ \hline
\multicolumn{4}{l}{\cellcolor[HTML]{EFEFEF}{\color[HTML]{000000} \textit{\textbf{(1) Zero-shot methods}}}} \\
Random & 50.5 & 33.7 & 40.3 \\
LAION-CLAP & \textbf{83.7} & 38.7 & \textbf{52.9} \\
MS-CLAP & \textbf{83.7} & 37.8 & 52.2 \\ \hline
\multicolumn{4}{l}{\cellcolor[HTML]{EFEFEF}\textit{\textbf{(2) Fine-tuning methods}}} \\
LAION-CLAP & 80.4 & 70.3 & 75.0 \\
MS-CLAP & \textbf{90.6} & \textbf{85.4} & \textbf{87.9} \\ \hline
\end{tabular}
\label{tab:classification_results}
\end{table}

\begin{table}[t]
\centering
\caption{The number of misclassified samples. For the samples that have hallucination labels, we count them for each type.}
\scalebox{0.8}{
\begin{tabular}{lcccc}
\hline
 & Not hallucinated & Type (A) & Type (B) & Type (C) \\ \hline
\multicolumn{5}{l}{\cellcolor[HTML]{EFEFEF}\textit{\textbf{(1) Zero-shot methods}}} \\
LAION-CLAP & 220 & 5 & 14 & 8 \\
MS-CLAP & 155 & 17 & 20 & 17 \\
\multicolumn{5}{l}{\cellcolor[HTML]{EFEFEF}\textit{\textbf{(2) Fine-tuning methods}}} \\
LAION-CLAP & 27 & 10 & 17 & 18 \\
MS-CLAP & 11 & 1 & 9 & 6 \\ \hline
\end{tabular}
}
\label{tab:mis_classified_samples}
\end{table}

\begin{figure*}[t]
  \centering
  \includegraphics[width=0.9\linewidth]{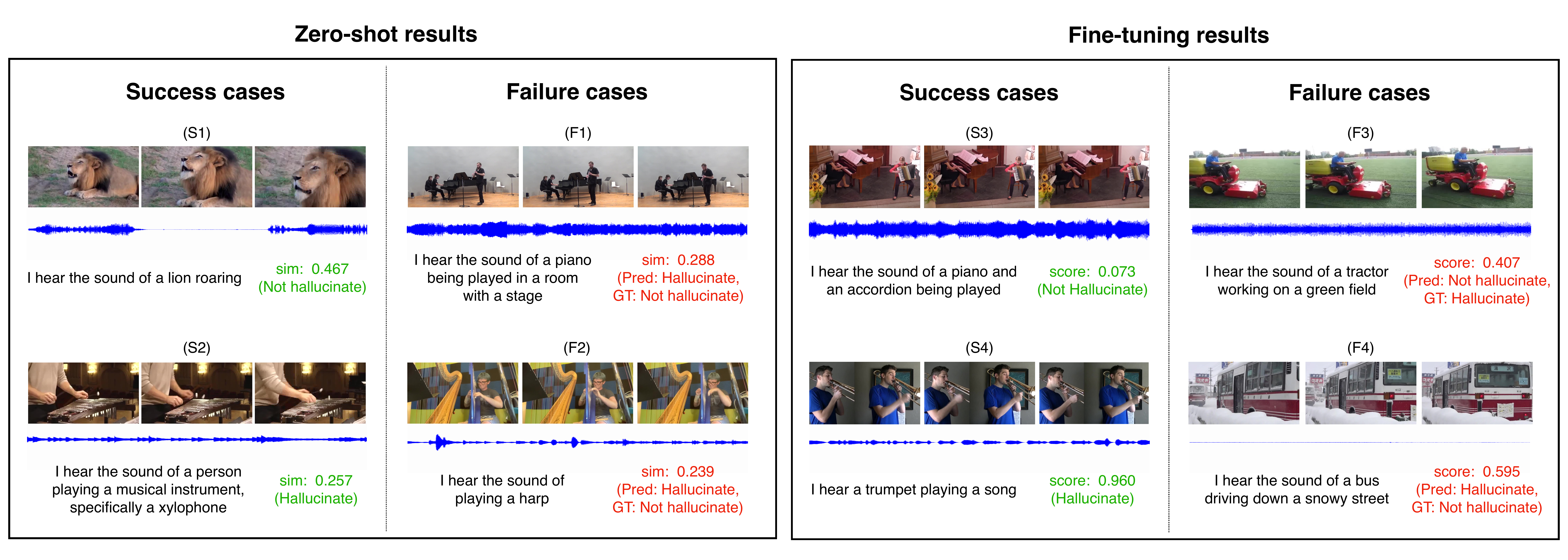}
  \caption{Success and failure cases in the zero-shot and fine-tuning methods.}
  \label{fig:success_failure_cases}
\end{figure*}

\subsection{Experimental settings}

As our dataset, we split the annotated 1,000 sentences with 400, 100, and 500 for training, validation, and testing. To evaluate the performance, we compute recall, precision, and F1 scores. As the pre-trained audio-text encoders, MS-CLAP \cite{CLAP2023} and LAION-CLAP \cite{laionclap2023} are used, which learns audio-text joint embedding space via contrastive learning as with CLIP \cite{radford2021icml}. For the zero-shot model, we set $\alpha$ to be $0.3$ and $0.45$ for MS-CLAP and LAION-CLAP. For training the fine-tuning model, we set the hidden size $d$ to $512$ and $256$ for MS-CLAP and LAION-CLAP. The batch size is set to be $32$. The adamW optimizer \cite{loshchilov2018decoupled} is used with the learning rate $0.001$ and $0.0001$ for MS-CLAP and LAION-CLAP.

\subsection{Quantitative evaluation}

\tabref{tab:classification_results} shows the classification performance in the zero-shot and fine-tuning settings. When comparing our zero-shot approach with the random baseline, our approach performs better than the random baseline by 10\% in F1. LAION-CLAP achieves slightly better than MS-CLAP by 0.7\%.

In terms of the fine-tuning setting, both models outperform the zero-shot methods, indicating the possibility of training the classifier to detect audio hallucinations.
When comparing backbones, we observe that MS-CLAP outperforms LAION-CLAP by +11\% in F1.
The cause of this discrepancy is unclear, but a potential factor is attributed to the nature of the text data in their training datasets.
MS-CLAP utilizes audio-sentence pairs from the web, whereas LAION-CLAP relies on augmented audio-sentence pairs generated through a keyword-to-caption template. This noisy augmentation may have a negative impact on the classification of audio hallucinations.

\tabref{tab:mis_classified_samples} shows the number of misclassified samples for each hallucination type. Note that ``Not hallucinated'' is counted if the model predicts ``Hallucinated'' but the ground-truth label is ``Not hallucinated.''
We observe that type (B) is the most frequent case among the hallucination types in general.
This indicates that the model fails to distinguish the sound of the actions of the object correctly. As shown in (B.2) of \figref{fig:hallucination_examples}, there are confusing instances in this category. Therefore, it is necessary to develop audio-text models that can distinguish the difference between object sounds and action-object sounds (e.g., a chainsaw and a chainsaw cutting wood).

\subsection{Qualitative evaluation}

\figref{fig:success_failure_cases} shows the success and failure cases in the zero-shot and fine-tuning settings.
The zero-shot model can correctly predict the single sound of animals, such as lions roaring in (S1). One of the merits of using audio-text models is to distinguish the sound of visually similar musical instruments; in (S2), the model can predict the audio hallucinations of xylophone (correctly, it is glockenspiel). However, it often fails to predict the mixed sound of musical instruments, such as (F1). In addition, it also fails to predict the sound of less-frequent instruments.

The fine-tuning model can answer such mixed sounds of musical instruments and is aware of the differences between musical instruments. For example, in (S3), the model successfully learns the sound of piano and accordion in (S3) and understands that the sound is not a trumpet in (S4) (correctly, it is trombone). In contrast, it fails to answer the confusing sound based on only the sound. For example, in (F3), the sound is not a tractor but a lawnmower. In (F4), the sound is driving down the snowy street. Visual information can aid the model in distinguishing these cases, making it imperative to devise a method for seamlessly integrating it into audio-text models.

\section{Conclusion}

Large audio-video language models sometimes fail to generate faithful audio descriptions because they ignore audio information and generate them solely based on visual information. This study referred to these as audio hallucinations and focused on analyzing them by attaching two types of annotations with the generated sentences: (1) audio hallucination labels indicating whether sentences are hallucinated and (2) hallucination types categorizing the nature of the hallucinated sentences. Our analysis suggested that 33.2\% sentences are hallucinated with different trends observed in nouns and verbs for each hallucination.
Based on this insight, we conducted the audio hallucination classification using pre-trained audio-text models in both zero-shot and fine-tuning settings.
The experimental results revealed that the zero-shot models outperform the random baseline in detecting hallucinations, and the fine-tuning models effectively identify them with notable accuracy.

\bibliographystyle{IEEEtran}
\bibliography{mybib}

\end{document}